\documentclass[aps,pra,notitlepage,nofootinbib,superscriptaddress,reprint]{revtex4-2}
\usepackage{graphicx}
\usepackage{multirow}
\usepackage[tbtags]{amsmath}
\usepackage{amssymb,amsthm,mathtools}
\usepackage{braket}
\usepackage{epstopdf,cancel,ulem}
\usepackage{epsf,latexsym,bbm,euscript}
\usepackage[colorlinks,bookmarks=false]{hyperref}

\usepackage{comment} 
\usepackage{color} 
\usepackage{cancel}

\begin{document}

\author{Carolyn E. Wood}
\author{Magdalena Zych}
\affiliation{Australian Research Council Centre for Engineered Quantum Systems, School of Mathematics and Physics, The University of Queensland, St Lucia, QLD 4072, Australia}

\title{Quantized mass-energy effects in an Unruh-DeWitt detector}

\begin{abstract}
A simple but powerful particle detector model consists of a two-level system coupled to a field, where the detected particles are the field excitations. This is known as the Unruh-DeWitt detector. Research using this model has often focused on either a completely classical description of the external degrees of freedom of the detector, or a full field-theoretic treatment, where the detector itself is described as a field. Recently there has been much interest in quantum aspects of the detector's center of mass---either described as moving in superposition along semiclassical trajectories, or dynamically evolving under a nonrelativistic Hamiltonian. However, the processes of interest---the absorption or emission of field particles---necessarily change the detector's rest mass by the amount of energy of the absorbed or emitted field quanta. Neither of the above models can capture such effects. Here we incorporate the quantization of the detector's mass-energy into the Unruh-DeWitt model. We show that internal energy changes due to emission or absorption are relevant even in the lowest energy limit. Specifically, corrections to transition rates due to the detector's mass changing cannot be ignored unless the entire center of mass dynamics is also ignored. Our results imply that one cannot have a consistent model of the Unruh-DeWitt detector as a massive particle without including the mass-energy equivalence. 
\end{abstract}
\maketitle

\section{Introduction}
The mass-energy equivalence of special relativity implies a universal coupling between the internal degrees of freedom (DoFs) of a composite particle, which describe its total internal energy, and the particle's center of mass (CoM)~\cite{Zych2011}. The resulting simple model of a particle with quantized mass-energy opened new avenues for exploring phenomena at the quantum-and-gravity interface, including gravitational decoherence and time dilation~\cite{Pikovski2015,Khandelwal2020,Smith2020,Grochowski2021}.  The coupling leads to relativistic effects even at low CoM energies, thus also opening new avenues for laboratory experiments~\cite{Rosi2017, Loriani2019}. The coupling also has consequences for the free propagation of composite particles~\cite{Pang2016} and their boosts~\cite{Paige2020}, and ultimately leads to the requirement of a class of states which describe semiclassical propagation of composite particles---which are different to the usually expected pure Gaussian state of the particle's CoM~\cite{WoodZych2021}.

Importantly, the universality of the coupling also leads to signatures of the mass-energy equivalence emerging where it would not have been expected. For example, failure to incorporate the mass-energy equivalence in the spontaneous emission process leads to an ``anomalous friction'' effect~\cite{Sonnleitner2017, SonnleitnerBarnett2018}. Mass-energy equivalence resolves this problem by allowing the particle's momentum to change during emission without changing its velocity.

The above naturally implies that the mass of a particle interacting with a field must be altered by the outgoing or incoming energy in any emission or absorption process. This in turn implies that the count made by a physical detector, of particles in some field state (when the detector is modeled as a physical system with its own dynamics, including the CoM), may also be affected if the mass-energy equivalence is not incorporated. The simplest model of such a detector, the Unruh-DeWitt (UDW) detector, is used widely to probe quantum fields in curved spacetime and to explore applications to relativistic quantum information. 

Traditionally, the UDW detector model consists of a pointlike particle with some internal degrees of freedom interacting with a quantum field along a classically prescribed worldline~\cite{DeWitt1979, BirrellDavies1984}. However, it was originally formulated through quantum field theory, beginning with the seminal explorations of the Unruh effect and Hawking radiation~\cite{Hawking1975, Unruh1976,GibbonsHawking1977}. See ref.~\cite{Crispino2008} for a review of applications of the Unruh effect, including particle detectors. The field-theoretic treatment continues to be explored to gain insights into vacuum entanglement, e.g.~\cite{Costa2009}, and recently in the context of quantum reference frames (associated with the detector)~\cite{GiacominiKempf2022}. Nonetheless, many recent works introduce additional features to both the detector and the field as needed to apply the model to more diverse or more realistic situations. These range from explorations of light-matter interactions~\cite{MartinMartinez2013} and the Casimir-Polder effect~\cite{Debski2021} to axion detection~\cite{Kanno2022} and neutrino oscillations~\cite{Torres2020}. 

Additionally, a formalism has been developed to describe an UDW detector following a superposition of different classical paths~\cite{Barbado2020, Foo2020UDWSup, Foo2021Thermality, Foo2021Harvesting}. This has further allowed the exploration of quantum causal structures through superpositions of detector-field interaction times~\cite{Henderson2020}, as well as through describing the dynamics of the UDW detector in genuine quantum spacetimes~\cite{Foo2021deSitterCat, Foo2021BHmass}.
  
Recently there has been interest in a yet different regime, motivated by operational considerations: An UDW detector described as a first-quantized particle with internal states coupled to a field, with the addition of external dynamics---i.e.~quantized CoM degrees of freedom. This has been explored, for example, to study the effects of the CoM motion~\cite{Guo2008} and coherent spreading of the CoM~\cite{Stritzelberger2020}, and to study the recoil of the detector's CoM due to interactions with field quanta~\cite{Sudhir2021}. An understanding of the dynamics of the UDW detector's CoM is also relevant for models of atom-light interactions. However, thus far the models either considered a nonrelativistic CoM of the detector, thus not including the mass-energy equivalence (a fact also noted in Ref.~\cite{Lopp2021}), or looked at formal expressions for transition amplitudes for a relativistic CoM.

In this work, we introduce relativistic corrections to the UDW detector model which account for the mass-energy equivalence---whereby the mass is different for each of the UDW detector's ground and excited states. As the detector changes mass only at the moment of excitation (or emission) one might expect no differences, compared to earlier models, for transitions between initial internal eigenstates. Additionally, as the resulting effects are relativistic corrections one might further expect that there is a sufficiently low energy regime where the mass-energy equivalence effects can be neglected.
We show that neither of these intuitions hold. Incorporating quantized mass-energy into the UDW model is necessary for a consistent description as it leads to corrections to the detector's emission and absorption rates that are on the same order as the effects of even the nonrelativistic CoM dynamics.

The article is structured as follows: We present our generalization of the UDW detector dynamics and derive the emission and absorption amplitudes for our model in Sec.~\ref{sec:transitions}. We then find the emission and absorption rates in Secs.~\ref{sec:emrates} and~\ref{sec:absrates}, and discuss the relevance of the relativistic corrections. We also discuss an interesting further subtlety related to the interpretation of the nonrelativistic notion of mass in light of the role of mass in the relativistic model. We compare the rates obtained in this work to those arising from a nonrelativistic CoM model~\cite{Stritzelberger2020} in Sec.~\ref{sec:compare}. Finally, we discuss implications of our results in Sec.~\ref{sec:discussion}. Throughout this work we set $\hbar=1$ but keep the speed of light $c$ in the expressions for ease of discussing relativistic vs nonrelativistic effects.

\section{Unruh-DeWitt detector with external dynamics and mass-energy equivalence}\label{sec:transitions}
The UDW detector is a model of a pointlike particle coupling linearly through the monopole operator $\vec{\mu}$ to a scalar field along the detector's predefined, classical worldline $\vec{x}(\tau)$. 
The associated Schr\"{o}dinger picture interaction Hamiltonian has the form
\begin{equation}\label{eq:OldInteractionH}
\hat{H}_{\mathrm{int}} = \lambda\hat{\mu}\hat{\phi}(\vec{x}),
\end{equation}
where $\lambda$ is the coupling strength, $\hat{\mu}$ the monopole moment operator, and $\hat{\phi}(\vec{x})$ the scalar field evaluated at the point $\vec{x}$ which represents the classical position of the detector.

In the case of a detector with two internal energy states, the monopole operator takes the form
\begin{equation}\label{eq:monopoleop}
\hat{\mu} = \ket{e}\bra{g} + \ket{g}\bra{e},
\end{equation}
where $\ket{g}$ and $\ket{e}$ are respectively the ground and excited energy states.

In the majority of recent works, the detector's free Hamiltonian included just the energy associated with its internal state~\cite{Crispino2008}. Here we also include the CoM energy required to accurately describe the dynamics of all relevant DoFs of the detector for its interaction with the field. 

The energy of a pointlike first-quantized particle with internal DoFs and in flat spacetime (Minkowski metric) reads $\sqrt{\hat{M}^2c^4+\hat{\vec{p}}^2c^2}$~\cite{Zych2011, Pikovski2015,SchwartzGiulini2019CQG,SchwartzGiulini2019PRA,Roura2020,Smith2020,WoodZych2021} where $\hat{M}c^2$ is the total rest mass-energy of the UDW detector. This form is required for detector models which aim to include both quantum and relativistic effects, see also Ref.~\cite{GiacominiKempf2022} for the corresponding second-quantized model. Here we are interested in the limit of low CoM energy---where first-quantization is accurately describing the system and where the detector's CoM is moving slowly---and investigate the impact of the mass-energy equivalence in this regime. 

The CoM energy of the composite particle in such a low-energy limit reduces to
\begin{equation}\label{eq:LowECoM}
\hat{M}c^2+\frac{\hat{\vec{p}}^2}{2\hat{M}}.
\end{equation}

Note, a fully nonrelativistic model of the detector would be governed by the Galilei symmetry group, however with the inclusion of internal dynamics through $\hat{M}$, the symmetry group of the resulting model in Eq.~\eqref{eq:LowECoM} is the central extension of the Galilei group~\cite{Giulini1996, ZychGreenberger2019}. Since this inclusion of the internal dynamics comes from relativistic mass-energy equivalence, we will hereafter refer to our model as the `semirelativistic' model, to distinguish it from a fully nonrelativistic model where the mass features only as a fixed parameter.

We restrict the model to just two internal states corresponding to the detector's ground $\ket{g}$ and excited $\ket{e}$ states, and introduce the following split
\begin{equation}\label{eq:MassOperator}
\begin{split}
\hat{M} &= m_g\ket{g}\bra{g} + m_e \ket{e}\bra{e}\\
&= m_g\ket{g}\bra{g} + \left(m_g + \frac{E}{c^2}\right) \ket{e}\bra{e}.
\end{split}
\end{equation}

Thus, $m_g, m_e$ are the masses associated with the ground and the excited state of the detector, respectively, and $E$ is the energy gap between them. Note that the first term in Eq.~\eqref{eq:LowECoM} gives the usual internal Hamiltonian of the UDW detector, while the second term is the kinetic energy with mass-energy incorporated. 

Note that the fully nonrelativistic limit of the detector's dynamics is obtained when $E/m_gc^2\rightarrow0$. This is in addition to the low-energy limit for the CoM already present in Eq.~\eqref{eq:LowECoM}, which requires  $\vec{p}^2\ll (m_{g(e)}c)^2$. The full nonrelativistic limit thus results in the kinetic term in Eq.~\eqref{eq:LowECoM} becoming $\hat p^2/2\hat M\rightarrow\hat p^2/2m_g\otimes\hat{\mathbb{I}}_{\mathrm{int}}$ where $\hat{\mathbb{I}}_{\mathrm{int}}$ is the identity operator on the internal states and where the first factor acts only on the CoM. Importantly, the rest energy term in Eq.~\eqref{eq:MassOperator} remains unchanged in this limit. 
The full nonrelativistic limit of Eq.~\eqref{eq:LowECoM}, dropping dynamically irrelevant term $m_gc^2 \hat{\mathbb{I}}$ arising from $\hat Mc^2$, reads
\begin{equation}\label{eq:non-relH0}
E\ket{e}\bra{e} + \frac{\vec{p}^2}{2m_g}.
\end{equation}
As expected in nonrelativistic physics, the above is simply  a sum of the internal energy (acting only on the internal DoF)  and the kinetic energy (acting only on the CoM) which features one mass parameter $m_g$ for all internal states. See refs~\cite{ZychGreenberger2019, ZychRudnicki2019} for detailed discussions of the nonrelativistic limit of dynamics for composite particles, which relate to the aforementioned underlying group structure.

The full free Hamiltonian in our model consists of the  energy of the detector, Eq.~\eqref{eq:LowECoM}, and the energy of the massless scalar field $ \int d^3 k~ c |\vec{k}| ~\hat{a}^{\dagger}_{\vec{k}}\hat{a}_{\vec{k}}$,  with $\hat{a}_{\vec{k}}, \hat{a}_{\vec{k}}^{\dagger}$ the creation and annihilation operators of the field quanta with three-momentum $\vec{k}$, satisfying the canonical commutation relation $[\hat{a}_{\vec{k}}, \hat{a}_{\vec{k}'}^{\dagger} ]=\delta^{(3)}(\vec{k} - \vec{k}')$. Dropping the term proportional to the identity from Eq.~\eqref{eq:MassOperator}, the total free Hamiltonian for our model reads
\begin{equation}\label{eq:MEEfreeH}
\hat{H}_0 = E \ket{e}\bra{e} +\frac{\hat{\vec{p}}^2}{2\hat{M}} + \int d^3 k~ c |\vec{k}| ~\hat{a}^{\dagger}_{\vec{k}}\hat{a}_{\vec{k}}.
\end{equation}

The detector couples to the field at the CoM position, described by the position operator $\hat{\vec{x}}$. Equation~\eqref{eq:OldInteractionH} is modified to include this  using the spectral expansion of the position operator
 \begin{equation}\label{eq:NewInteractionH}
\hat{H}_{\mathrm{int}} = \lambda\int d^3 x \ket{\vec{x}}\bra{\vec{x}} \otimes \hat{\mu} \otimes \hat{\phi}(\vec{x}),
\end{equation}
with CoM position eigenvalues $\vec{x}$, and eigenstates $\ket{\vec{x}}$, and with the field operator defined as
\begin{equation}\label{eq:fieldop}
\hat{\phi}(\vec{x}) = \int \frac{d^3k}{(2\pi)^{3/2}}\sqrt{\frac{c^2}{2|\vec{k}|}} \left[e^{i\vec{k}\vec{x}}\hat{a}_{\vec{k}} + e^{-i\vec{k}\vec{x}}\hat{a}^{\dagger}_{\vec{k}} \right],
\end{equation}
where normalization of the field has been chosen such that $\lambda$ remains dimensionless.

To find the transition amplitude for the system, we work in the interaction picture and perform a perturbative expansion. The interaction picture form of the interaction Hamiltonian is obtained from $e^{i H_0 t}\ket{\vec{x}}\bra{\vec{x}}\hat{\mu}\hat{\Phi}(\vec{x}) e^{-i H_0 t}$. 
Note that  due to the mass-energy equivalence, in the kinetic term of \eqref{eq:LowECoM} there is coupling between the CoM and internal DoFs of the detector and so the time evolution of $\hat{\mu}$ and $\ket{\vec x}\bra{\vec x}$ has to be computed jointly. In summary, Eqs~\eqref{eq:MEEfreeH} and \eqref{eq:NewInteractionH} result in the following interaction-picture Hamiltonian
\begin{multline}\label{eq:InteractionPicInteractionH}
\hat{H}_{\mathrm{int}}(t) = \lambda \int \hat{\phi}(\vec{x},t) \Bigg[ e^{it\left[\frac{\hat{\vec{p}}^2}{2m_e} + E \right]}\ket{e}\bra{g} \\
 \ket{\vec{x}(t)}\bra{\vec{x}(t)} e^{-it \frac{\hat{\vec{p}}^2}{2m_g}} + \mathrm{H.c.} \Bigg]d^3x,
\end{multline}
where the time-dependent position eigenstates can be expressed as
\begin{equation}
\ket{\vec{x}(t)}\ket{i} = \int \frac{d^3 p}{(2\pi)^{3/2}} e^{-i\vec{p}\vec{x}+it\frac{\vec{p}^2}{2}\left(\frac{1}{m_g}\ket{g}\bra{g}+\frac{1}{m_e}\ket{e}\bra{e}\right)}\ket{\vec{p}}\ket{i}.
\end{equation}
where $\ket{i}$ is an arbitrary state of the internal DoF.

Note that in the fully nonrelativistic limit of the CoM we recover the model of the UDW detector with mass as a nondynamical parameter and fixed internal energy gap $E$ which is used in ref.~\cite{Stritzelberger2020}. The question of whether the ground state mass-energy $m_g$ or the excited state’s $m_e$ becomes the mass parameter of the nonrelativistic model is discussed in Sec. \ref{sec:compare}.

Transition amplitudes between any pair of mutually orthogonal states, to first order perturbation in the coupling strength $\lambda$, are given by
\begin{equation}\label{eq:amplitude}
\mathcal{A} = -i\bra{\Psi_f}\int^{t_f}_{t_i} dt ~\hat{H}_{\mathrm{int}}(t)\ket{\Psi_i}
\end{equation}
with $\ket{\Psi_i}$ and $\ket{\Psi_f}$ the initial and final states of the system, respectively.

We first look at the emission process and thus consider the following states
\begin{equation}\label{eq:initandfinalemiss}
\ket{\Psi_i} = \ket{\psi_0} \otimes \ket{e} \otimes \ket{0},~~\ket{\Psi_f} = \ket{\vec{p}'} \otimes \ket{g} \otimes \hat{a}^{\dagger}_{\vec{k}'}\ket{0},
\end{equation}
describing the detector beginning in the excited state, with an arbitrary CoM state denoted by $\ket{\psi_0}$, and where the field is initially in the Minkowski vacuum state. The final state is then taken as the ground state of the detector with some CoM momentum $\vec{p}'$, and a one-particle excitation in the field. We will later sum over the final states of the detector and the field to derive the total emission rate, discussed in Sec.~\ref{sec:emrates}.

The states in Eq.~\eqref{eq:initandfinalemiss}, incorporated into Eq.~\eqref{eq:amplitude}, result in the following emission amplitude:
\begin{multline}\label{eq:emissionamp}
\mathcal{A}^{\mathrm{em}}_{\mathrm{semirel.}} = -\frac{i\lambda}{(2\pi)^{9/2}}\sqrt{\frac{c^2}{2|\vec{k}|}}\psi_0(\vec{p}'+\vec{k})\\
\times \int^{t_f}_{t_i} dt ~e^{it(\frac{\vec{p}'^2}{2m_g}-\frac{(\vec{p}'+\vec{k})^2}{2m_e}-E + c|\vec{k}|)}.
\end{multline}

Both masses $m_g$ and $m_e$ are present in the time integral and in general affect the amplitude. This integral (with bounds extended to infinity) expresses  the energy conservation in the process in which the internal state and thus  also the mass-energy of the detector changes. This clarifies why the amplitude in general depends on the masses of both internal states. This also indicates that in general the mass-energy equivalence will have consequences for observable effects, which we  quantitatively assess in the next section. Note that the difference between the masses disappears from Eq.~\eqref{eq:emissionamp}, for arbitrary final states of the field, only when the initial momentum of the detector is strictly zero. That is, when $\psi_0(\vec{p})\propto\delta^{(3)}(\vec{p})$, in which case the amplitude depends only on the ground state mass $m_g$. 

We now look at the absorption amplitude, where the detector begins in the ground state. 
In the present case, as the absorption or emission of a field particle affects the initial and final mass of the detector, the two processes exhibit differences that are not accessible in an UDW detector with a preassigned trajectory nor in a model with fully nonrelativistic CoM. The initial and final states of all DoFs, in analogy to the emission case, read:
\begin{equation}\label{eq:initandfinalabs}
\ket{\Psi_i} = \ket{\psi_0} \otimes \ket{g} \otimes \hat{a}^{\dagger}_{\vec{k}'}\ket{0},~~\ket{\Psi_f} = \ket{\vec{p}'} \otimes \ket{e} \otimes \ket{0}.
\end{equation}
Note that for comparison between the two processes we take the same initial state $\ket{\psi_0}$ for the CoM as in Eq.~\eqref{eq:initandfinalemiss}.
The resulting amplitude for the absorption process is
\begin{multline}\label{eq:absorptionamp}
\mathcal{A}^{\mathrm{abs}}_{\mathrm{semirel.}}= -\frac{i\lambda}{(2\pi)^{9/2}} \sqrt{\frac{c^2}{2|\vec{k}|}}~e^{-it_f\left(\frac{\vec{p}'^2}{2m_e} + E\right)} ~\psi_0(\vec{p}'-\vec{k})\\
\times \int^{t_f}_{t_i} dt ~e^{it\left(\frac{\vec{p}'^2}{2m_e}-\frac{(\vec{p}'-\vec{k})^2}{2m_g}+E-c|\vec{k}|\right)}.
\end{multline} 
Compared to the emission case, the signs of $E$ and $k$, as well as the masses $m_g, m_e$ have swapped places, as expected.  We thus again see both masses in the time integral, contributing to conservation of energy (for $t_i,t_f$ taken to infinity). As in the emission case, only one of the masses plays a role in the final amplitude for the initial CoM state being  $\psi_0(\vec{p})\propto\delta^{(3)}(\vec{p})$, however, for the absorption process this is the excited state mass $m_e$. 
In general, the two masses contribute differently to the two amplitudes (true also for other CoM states), which has consequences for the interpretation of the nonrelativistic mass parameter of the detector, further discussed in the next two sections.
The rate of absorption corresponding to Eq.~\eqref{eq:absorptionamp} is explored in Section~\ref{sec:absrates}.

\section{Mass-energy effect on detector emission and absorption rates}\label{sec:rates}

\subsection{Emission rate}\label{sec:emrates}
The total emission probability corresponding to the amplitude Eq.~\eqref{eq:emissionamp} is obtained by summing over the final momenta of the CoM and of the field. The corresponding transition rate is then obtained by taking its time derivative, and then taking the limits of the time integral to infinity. The resulting transition rate for the emission process is
\begin{equation}\label{eq:fullemrate}
\mathcal{R}^{\mathrm{em}}_{\mathrm{semirel.}} = \frac{\lambda^2 c^2 m_g}{4\pi} \int d^3p ~\left|\psi_0(\vec{p})\right|^2 \mathcal{T}^{\mathrm{em}}(p),
\end{equation}
where $p\equiv |\vec{p}|$ and
\begin{equation}
\begin{split}\label{eq:myemtemplate}
\mathcal{T}^{\mathrm{em}}_{\mathrm{semirel.}}(p)&:= 2 - \frac{1}{p} \sqrt{\frac{p^2 m_g}{m_e}+2pm_gc+m_g^2c^2+2Em_g}\\
&\qquad +\frac{1}{p}\sqrt{\frac{p^2 m_g}{m_e}-2pm_gc+m_g^2c^2+2Em_g}.
\end{split}
\end{equation}

We refer to $\mathcal{T}(p)$ as a ``template function'' following the terminology used in Ref.~\cite{Stritzelberger2020}. In the nonrelativistic limit for the detector's CoM, discussed in the previous section, we have $m_g = m_e$, and Eq.~\eqref{eq:myemtemplate} reduces to the expression found in that paper. In the infinite mass limit (while keeping $E$ finite), for  any normalized initial CoM state $\psi_0$, 
Eqs~\eqref{eq:fullemrate}, \eqref{eq:myemtemplate} reduce to the ``classical'' case, i.e.,~a detector on a fixed classical trajectory:
\begin{equation}\label{eq:classrate}
\mathcal{R}^{\mathrm{em}}_{\mathrm{class.}} = \frac{\lambda^2}{2\pi} E.
\end{equation} 

Because our Hamiltonian is valid for low CoM energies, we expand around small CoM  momentum, specifically, about $p/m_gc=0$. Note that this is also the regime where we expect the least impact of the mass-energy equivalence on the final result, as discussed below Eqs~\eqref{eq:emissionamp}, \eqref{eq:absorptionamp}. The expansion results in a template function
\begin{equation}
\mathcal{T}^{\mathrm{em}}_{\mathrm{semirel.}}(p) = 2-\frac{2}{\sqrt{1 +\frac{2E}{m_gc^2}}}+\frac{p^2 (c^2(m_g - m_e)+2E)}{c^4 m_g^2 m_e \left(1+\frac{2E}{m_gc^2}\right)^{\frac{5}{2}}}.
\end{equation}

We now take for the CoM state a three-dimensional Gaussian wave packet centered at $\vec{p}=0$, with position-space wave function 
\begin{equation}\label{eq:testGaussian}
\psi_0(\vec{x})= \left(\frac{2}{\pi L^2}\right)^{3/4} e^{-\frac{|\vec{x}-\vec{x}_0|^2}{L^2}},
\end{equation}
where $L$ is the initial (spatial) spread of the wave packet, centered in space at $\vec{x} = \vec{x}_0$. The resulting emission rate reads
\begin{multline}\label{eq:tayloremrate}
\mathcal{R}^{em}_{\mathrm{semirel.}}=\frac{\lambda^2 c^2 m_g}{2\pi}\\ \Bigg[1 -\frac{1}{\sqrt{1+\frac{2E}{m_g c^2}}}+\frac{3~(c^2(m_g-m_e)+2E)}{2L^2c^4m_g^2 m_e \left(1+\frac{2E}{m_gc^2}\right)^{\frac{5}{2}}}\Bigg].
\end{multline}
We note that the above holds provided that  $L>1/m_gc$, i.e.,~the localization of the wave packet, is larger than the Compton wavelength, as localization below this scale can lead to unphysical results (due to the breakdown of 1st-quantization). Equivalently, the expression holds when the initial momentum spread is small, $L_p:=1/L< m_gc$,  so that the initial state is consistent with the low-energy limit of the detector's CoM dynamics.

For $m_g = m_e \equiv M$ in Eq.~\eqref{eq:tayloremrate}, we again recover the result of ref.~\cite{Stritzelberger2020} for the nonrelativistic CoM:
\begin{multline}\label{eq:SKemrate}
\mathcal{R}^{\mathrm{em}}_{\mathrm{nonrel.}} =\frac{ \lambda^2 c^2 M }{2\pi} \\\Bigg[1-\frac{1}{\sqrt{1+\frac{2E}{Mc^2}}}+\frac{3 E}{L^2 c^4 M^3 \left(1+\frac{2E}{Mc^2}\right)^{\frac{5}{2}}}\Bigg],
\end{multline}
Note that this limit cannot be achieved by simply setting $E$ to zero in the final rate, Eq.~\eqref{eq:tayloremrate}, since  $E$ plays two roles in any relativistic model of an emitter/absorber--- it is the ``gap'' between masses associated with different internal states in the free Hamiltonian, and  the  energy gap between internal states in the interaction Hamiltonian. In prior studies, the UDW detector had no mass gap, but a finite internal energy gap, and so to recover such models it is more convenient to keep the general notation for the masses $m_g, m_e$  (despite the fact that their difference is simply $E/c^2$). 

From the physics perspective,  in these prior models interaction with the field would only alter the linear momentum of the CoM of the detector but not its mass. As discussed above, by incorporating the mass-energy equivalence, known in atomic physics as the R{\"o}ntgen term~\cite{Lopp2021} or the mass-defect~\cite{Yudin:2018, Haustein2019, Hammerer2022Abinitio}, our model can take into account the energy change of the CoM arising also due to the change in the mass of the detector.

Importantly, this mass change cannot be neglected even if one is looking only at transitions between internal states, as we do here. Indeed, to lowest order in $E/m_gc^2$, our Eq.~\eqref{eq:tayloremrate} reads $\mathcal{R}^{\mathrm{em}}_{\mathrm{semirel.}} \sim  E (1+\frac{3}{2}\frac{1}{ (Lm_gc)^2 }) $ while the nonrelativistic rate Eq.~\eqref{eq:SKemrate} reads $\mathcal{R}^{\mathrm{em}}_{\mathrm{nonrel.}}\sim  E (1+3\frac{1}{(Lm_gc)^2 })$.  This is the same for both choices of  $M=m_g$ and $M=m_e=m_g+E/c^2$ in Eq.~\eqref{eq:SKemrate}. Thus, the mass-energy effects cannot be ignored even at lowest order; they are only irrelevant in the ``classical'' limit where the dynamics of the CoM is entirely ignored, Eq.~\eqref{eq:classrate}. 

We further compare the full emission rates Eqs \eqref{eq:tayloremrate} and ~\eqref{eq:SKemrate} numerically in Sec.~\ref{sec:compare}.

\subsection{Absorption rate}\label{sec:absrates}

Following the same steps as for the emission process, the absorption rate arising from~Eq.~\eqref{eq:absorptionamp} is 
\begin{equation}\label{eq:fullabsrate}
\mathcal{R}^{\mathrm{abs}}_{\mathrm{semirel.}}= \frac{\lambda^2 c^2 m_e}{2\pi} \int d^3 p  \left|\psi_0(\vec{p})\right|^2 \mathcal{T}^{\mathrm{abs}}(p),
\end{equation}
with template function
\begin{multline}\label{eq:myabstemplate}
\mathcal{T}^{\mathrm{abs}}_{\mathrm{semirel.}}(p):= \frac{1}{p}\sqrt{\frac{p^2 m_e}{m_g}+2pm_e c+m_e^2c^2-2m_eE}\\
 -\frac{1}{p}\sqrt{\frac{p^2 m_e}{m_g}-2pm_e c+m_e^2c^2-2m_eE}.
\end{multline}
The expression for the rate formally requires us to introduce a cutoff $K$ in the momentum integral (i.e.~$p<K$) and to restrict the energy gap to $E/m_gc^2<(\frac{K}{m_gc}-1)^2$. We treat the width $L_p$ of the Gaussian state as an effective cutoff $K$ henceforth.
As anticipated, in the absorption case above the masses $m_g$ and $m_e$ have swapped compared to the emission case, Eq.~\eqref{eq:myemtemplate}, and the sign of the energy (in $2m_eE$) has changed.

Taking again the expansion about $p/m_gc=0$, and using the Eq.~\eqref{eq:testGaussian} Gaussian, we find the absorption rate
\begin{multline}\label{eq:plottedabsrate}
\mathcal{R}^{\mathrm{abs}}_{\mathrm{semirel.}} = \frac{ \lambda^2 c^2 m_e}{\pi} \\ \Bigg[\frac{1}{\sqrt{1-\frac{2 E}{m_e c^2}}}+\frac{3\left(c^2(m_g-m_e)+2E\right)}{2L^2 c^4 m_e^2 m_g \left(1-\frac{2E}{m_ec^2}\right)^{\frac{5}{2}}}\Bigg].
\end{multline}

As in the emission case we can take $m_g = m_e \equiv M$ in Eq.~\eqref{eq:plottedabsrate}, to obtain the result for a nonrelativistic CoM:
\begin{multline}\label{eq:SKabsrate}
\mathcal{R}^{\mathrm{abs}}_{\mathrm{nonrel.}} =\frac{ \lambda^2 c^2 M}{\pi}\\ \Bigg[\frac{1}{\sqrt{1-\frac{2 E}{M c^2}}}+\frac{3 E}{L^2 c^4M^3 \left(1-\frac{2E}{Mc^2}\right)^{\frac{5}{2}}}\Bigg].
\end{multline}

We now look again at the rates to lowest order in mass-energy corrections $E/m_gc^2$. Our result reads $\mathcal{R}^{\mathrm{abs}}_{\mathrm{semirel.}}\sim 1+2E(1+\frac{3}{4}\frac{1}{(Lm_gc)^2}) $, while the nonrelativistic rate is $\mathcal{R}^{\mathrm{abs}}_{\mathrm{nonrel.}}\sim 1+2E(1+\frac{3}{2}\frac{1}{(Lm_gc)^2}) $ for $M=m_e$ and reads $\mathcal{R}^{\mathrm{abs}}_{\mathrm{non-rel.}}\sim 1+E(1+3\frac{1}{(Lm_gc)^2}) $ for $M=m_g$. We again see that mass-energy corrections cannot be ignored. However, we also see that with the choice of $M=m_e$ the nonrelativistic expression better approximates the relativistic result---a conclusion which also holds for full rates, which we will further discuss based on the figures in the next section.

Finally, we note that for the absorption process, the classical limit is undefined in the sense that the above rates diverge in the limit of large mass. We thus will not be comparing the absorption rates against such a classical limit of a detector.

\subsection{Comparing to prior models}\label{sec:compare}

In previous models that looked at a first-quantized UDW detector, or in a related model of a dipole interacting with an optical field, the CoM of the detector/dipole was either pre-fixed (nondynamical), or dynamical but nonrelativistic, having a fixed mass parameter $M$,~e.g.~\cite{Guo2008}. In this work, we incorporate the relativistic correction to the CoM which allows the mass of the detector to change upon the exchange of energy with the field (where in an emission process, the mass is decreasing, while in an absorption process it is increasing).
Therefore, in order to compare our results with nonrelativistic models, we must interpret how the mass-energies $m_g, m_e$ relate to the nonrelativistic mass parameter $M$. 

One choice is to equate $M$ with the ground state mass $m_g$. This choice naturally arises when looking at a nonrelativistic limit for the free Hamiltonian discussed in Sec.~\ref{sec:transitions}. It is worthwhile to note that this itself arises from taking the split of $\hat M$ as in Eq.~\eqref{eq:MassOperator}, and so defining internal energy such that the internal ground state $\ket g$ has (internal) energy zero and the energy of the excited state $\ket e$ is non-negative. 

Another choice is to equate the nonrelativistic mass $M$ with the mass-energy which allows the relativistic and nonrelativistic dynamics of the full interacting system to align more closely (in the limit of small difference between $m_e, m_g$) . Surprisingly, this approach dictates different choices for $M$ for the emission and for the absorption processes. It is only in the full nonrelativistic limit, where $m_e\equiv m_g$, that the ambiguity is resolved, as there is strictly only one mass parameter. 
In order to understand which choice for the value of the non-relativistic mass $M$ best approximates the relativistic corrections---which albeit small are nonetheless finite---we explore both choices and discuss them further in this section.

Before presenting plots as illustrations of our analytical results, we reiterate that the form of the absorption rates Eqs~\eqref{eq:plottedabsrate}, \eqref{eq:SKabsrate} requires the parameters to be constrained so that $2E<m_ec^2$ and $2E<Mc^2$, respectively. As a result, we have chosen to present both the absorption and the emission rates' dependence on $E$ in this regime for consistency, even though the latter does not technically require this restriction.

\subsubsection{Emission rates compared}

\begin{figure}
\vspace{0.5 cm}
\centering
	\includegraphics[width=0.5\textwidth]{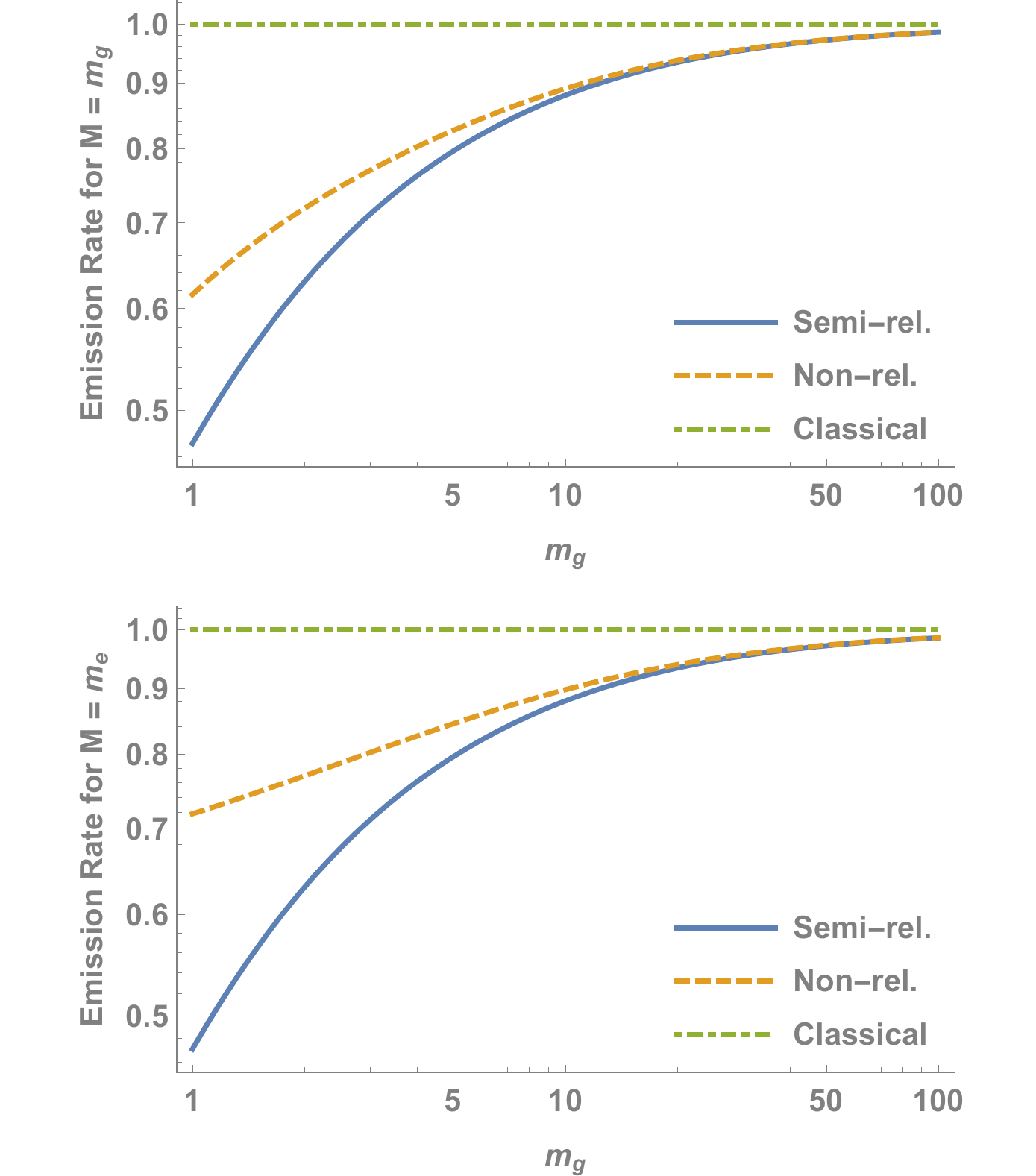}
	\caption{Emission rates as a function of mass for the UDW detector: With classical (fixed) CoM---the green, dashed-dotted line; with dynamical CoM including relativistic mass-energy corrections (our semirelativistic model)---the solid blue line; and with nonrelativistic CoM---dashed yellow line.  Top: for the nonrelativistic mass set to $M=m_g$. Bottom: for $M=m_e$. The classical UDW case can be seen as the infinite-mass limit of both models (ours and the nonrelativistic one) and for both choices of the mass, as discussed in the main text. All quantities are in units of the energy gap (setting $c=1$), thus we plot starting from $m_g/E=1$ and we use $L\cdot E=1$ which guarantees that $L>1/m_g$. The rates are additionally scaled by $(\lambda^2/ 2\pi)^{-1}$ setting the classical rate to $1$. Finally, both axes are on a log scale.}
	\label{fig:RvsM_ClassComp}
\end{figure}

In Fig.~\ref{fig:RvsM_ClassComp} the emission rate for the classical (infinite mass) case, Eq.~\eqref{eq:classrate}, is compared directly against our result~\eqref{eq:tayloremrate} (the ``semirelativistic'' model) and the rate for the nonrelativistic CoM model~\eqref{eq:SKemrate}, for each of the choices of $M \equiv m_g$ (top plot) and $M \equiv m_e$ (bottom plot). All quantities are presented in units of the energy gap $E$ (setting $c=1$), with the rates scaled by $(\lambda^2/2\pi)^{-1}$, resulting in a value of 1 for the classical rate.

The convergence of the other two rates to the classical case previously mentioned in Sec.~\ref{sec:emrates} is evident here. Our model is exactly the same in the top and bottom plots, while the curve for the non-relativistic model changes depending on whether its mass $M$ is identified with $m_g$ or $m_e$. The form of the non-relativistic curves thus differs from the top to the bottom plot because of the different mass used in the non-relativistic emission rate, Eq.~\eqref{eq:SKemrate}. In the bottom plot this is $M = m_e = m_g + E$, while in the top plot $M=m_g$. This will continue to underpin the behaviour of the non-relativistic model in subsequent figures below.

\begin{figure*}
\vspace{0.5 cm}
\centering
	\includegraphics[width=\textwidth]{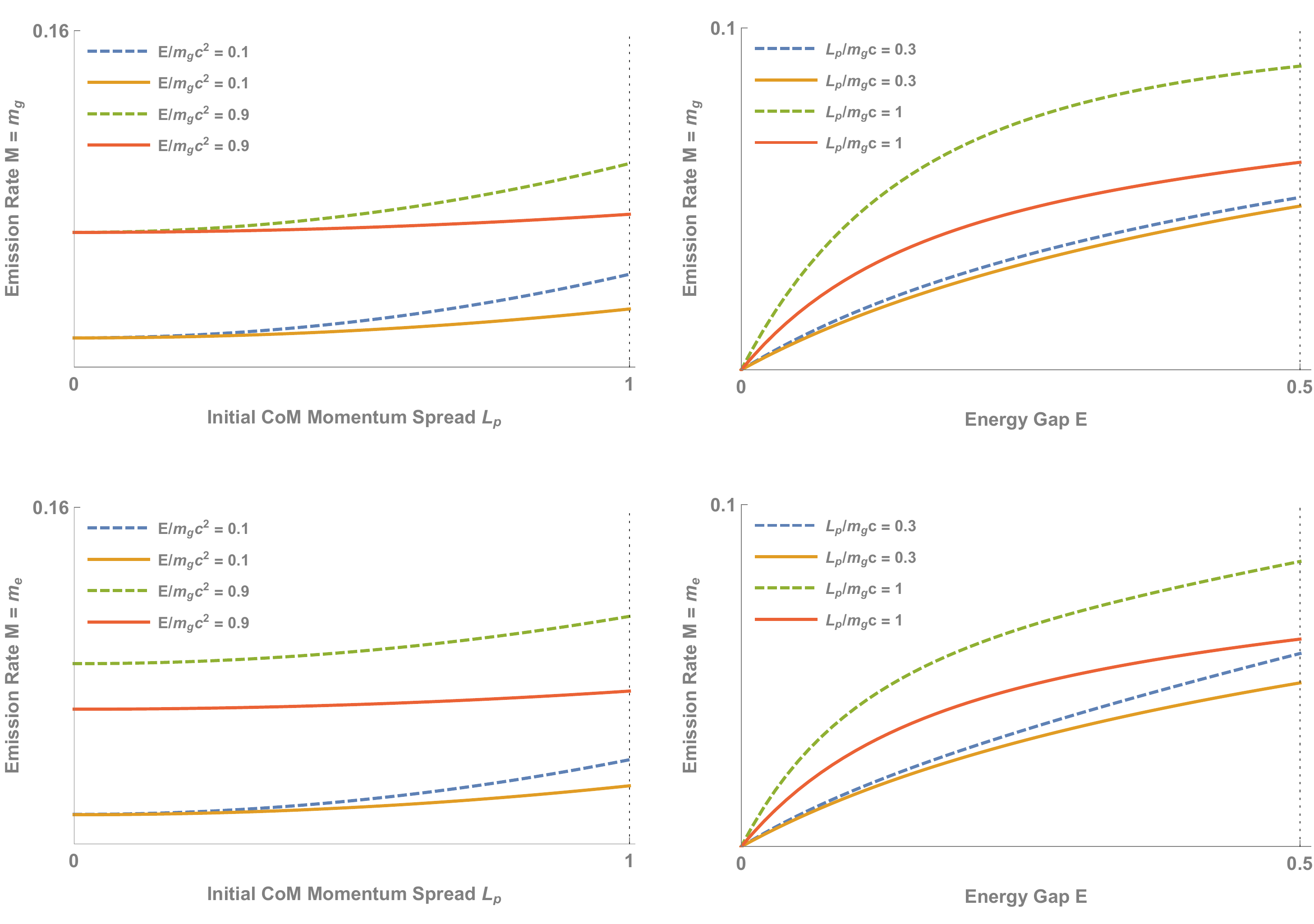}
	\caption{Emission rates $\mathcal{R}^{\mathrm{em}}_{\mathrm{semirel.}}$ for our model of the UDW detector with relativistic corrections from the mass-energy equivalence (solid lines) and for the model with nonrelativistic CoM (dashed lines). Left column: emission rate dependence on the initial CoM spread in momentum, $L_p$, for the two choices of the nonrelativistic model's mass $M=m_g(m_e)$ top (bottom) panels. Right column: emission rate dependence on the energy gap $E$ between ground and excited states, top/bottom panels as in the left column. All quantities are in units set by $m_g$, that is: rates in units of the Compton frequency $1/m_gc^2$, $L_p$ in units of $m_gc$, and $E$ in units of $m_gc^2$. The vertical dotted line at $L_p=1\cdot (m_gc)$ marks the limit of the applicability of the first-quantized model.
}
	\label{fig:RvsLandRvsEem_grid}
\end{figure*}

We now further explore the behavior of the emission rate of our model compared to that of the nonrelativistic CoM model for both choices of the nonrelativistic model's mass. We plot Eq.~\eqref{eq:tayloremrate} and Eq.~\eqref{eq:SKemrate} for these two choices in Figure~\ref{fig:RvsLandRvsEem_grid}---the $M\equiv m_g$ case on the top row, and the $M\equiv m_e$ case on the bottom row. The  plots in the left column show the emission rate dependence on the initial spread in the CoM momentum $L_p=1/L$; the plots in the right column show the rate dependence on the energy gap $E$. We stress that in all these plots, as in Fig.~\ref{fig:RvsM_ClassComp}, our model is fixed while the nonrelativistic model's $M$ is different between the top and the bottom rows.

First we find that in general the inclusion of mass-energy equivalence suppresses the emission rate compared to the nonrelativistic CoM model. For increasing initial spread in momentum, $L_p$, the rates arising from the two models diverge as expected. That is, increasing $L_p$ means the higher CoM momenta are contributing significantly to the initial state and so relativistic corrections due to ${p}^2/m_g \neq {p}^2/m_e$  become more relevant. The difference is greater for the $M\equiv m_e$ case where it also increases with the energy gap.

Similar conclusions could be drawn for emission rate dependence on the energy gap $E$ based on the regime plotted. However, note that when looking at the limit of large energy $E\gg m_gc^2$ (or $E\gg Mc^2$), our emission rate converges to $\sim m_g$ while the nonrelativistic rate converges to $\sim M$. Hence the two converge to each other for $M=m_g$. This can be understood by looking at the template function Eq~\eqref{eq:myemtemplate} which approaches a constant value of 2 for large $E$, independently of whether $m_g\neq m_e$ or $m_g=m_e\equiv M$ as in the nonrelativistic case. Physically, this is the limit where the CoM dynamics are essentially completely suppressed by the rest energy. The first term in Eq.~\eqref{eq:LowECoM} then dominates over the second term, and  the dynamics of the two models become effectively the same. 
 
As we have mentioned earlier in this section, the choice of $M=m_g$ or $M=m_e$ for any finite $E$ results in different emission rates predicted by the nonrelativistic model. Considering that the ``correct'' choice here is the one which arises in the nonrelativistic limit of small $L_p$, this favors the choice of $M=m_g$. This is visible in all the emission rate plots in Fig.~\ref{fig:RvsLandRvsEem_grid}, most notably from the left-hand column (where the rates are plotted as a function of $L_p$ for these two choices of $M$).

\begin{figure*}[ht!]
\vspace{0.5 cm}
\centering
	\includegraphics[width=\textwidth]{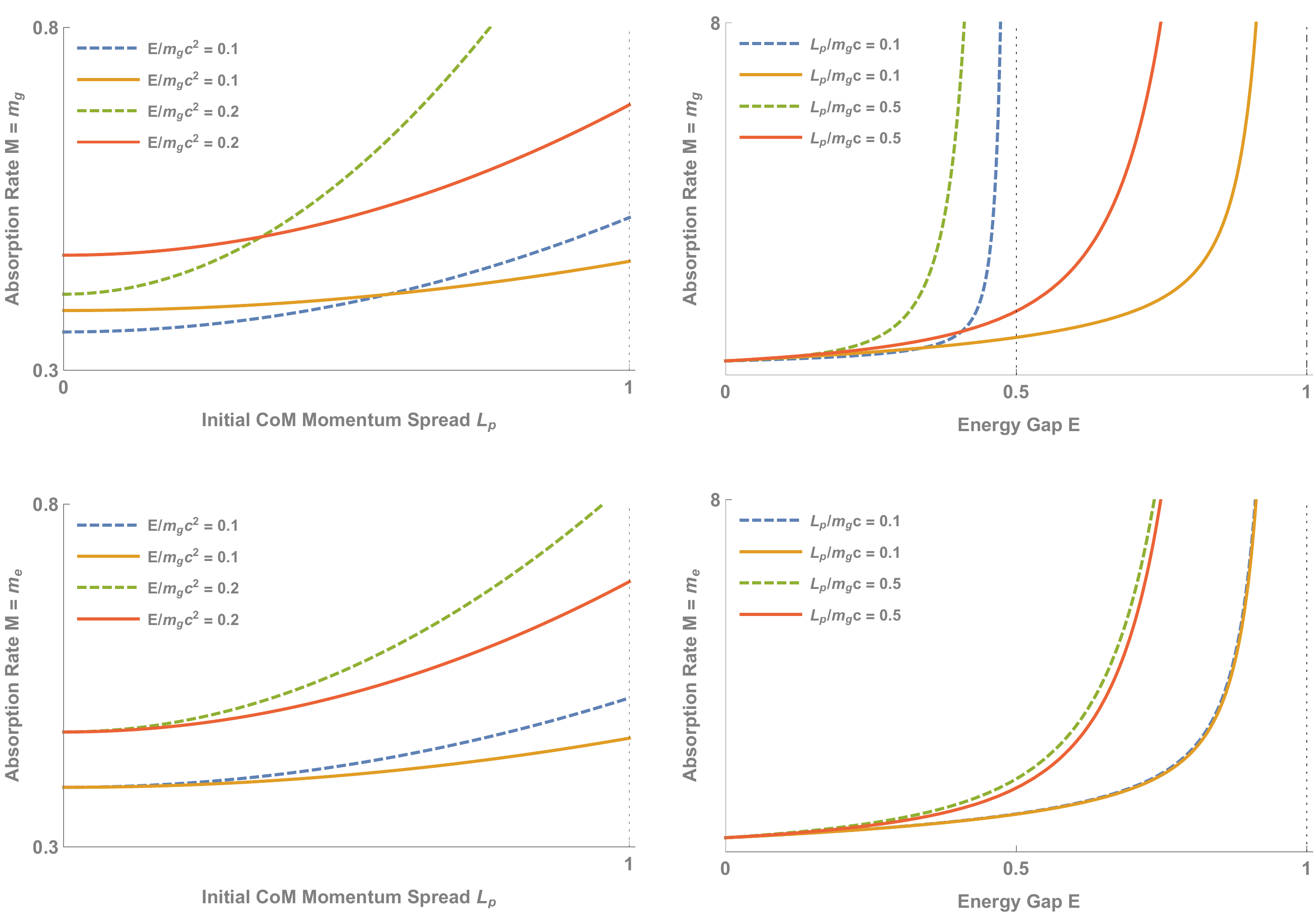}
\caption{Absorption rates $\mathcal{R}^{\mathrm{abs}}_{\mathrm{semirel.}}$ for our model of the UDW detector with relativistic corrections from the mass-energy equivalence (solid lines) and for the model with nonrelativistic CoM (dashed lines). Left column: absorption rate dependence on the initial momentum spread $L_p$. Right column: absorption rate dependence on the energy gap $E$. Top (bottom) panels in both columns correspond to the two choices for the nonrelativistic model's mass $M=m_g(m_e)$. Units of all quantities set by $m_g$: Rates are in the units of the Compton frequency $1/m_gc^2$, $L_p$ is in units of $m_gc$, and $E$ is in units of $m_gc^2$. The vertical dotted line at $L_p=1$ marks the limit of the applicability of the first-quantized model; the vertical lines in the right-column mark asymptotes of the absorption rates which are at $2E =m_ec^2$ (equivalent to $E=m_gc^2$) for our model and at $2E=Mc^2$ for the nonrelativistic model.}
	\label{fig:RvsLandRvsEabs_grid}
\end{figure*}

\subsubsection{Absorption rates compared}
We now turn to the absorption rate. 
As mentioned in the preceding section, absorption rates for both models considered here diverge in the large-mass limit. We thus move on to comparing the absorption rate of our model to that of the UDW detector with nonrelativistic CoM and we again do this for both choices for the nonrelativistic mass. Figure~\ref{fig:RvsLandRvsEabs_grid} thus plots the absorption rate for our model, Eq.~\eqref{eq:plottedabsrate}, and the equivalent for the UDW detector with nonrelativistic CoM, Eq.~\eqref{eq:SKabsrate}. As for the emission rate, the left-hand column shows the rates as a function of the initial CoM spread $L_p$ and the right-hand column the rates as a function of the energy gap. For the top row plots, the mass in the nonrelativistic model is set $M= m_g$ and in the bottom row it is set $M= m_e$. The asymptotes in the right column plots are at $2E=Mc^2$ for the nonrelativistic model and at $2E=m_ec^2$ (equivalently, at $ E=m_gc^2$, due to $m_e=m_g+E/c^2$) for our model. This is explicitly visible from the Eqs~\eqref{eq:plottedabsrate}, \eqref{eq:SKabsrate}.

Note that in this case, when looking at the limit of small momenta, small $L_p$, the two rates converge if we identify the nonrelativistic mass $M$ with the excited state mass $m_e$. This is the opposite of the emission case. For this choice (and thus for the bottom-row plots) we again see that the inclusion of relativistic corrections overall suppresses the rate compared to the nonrelativistic model.   

\section{Discussion}\label{sec:discussion}
We derived transition amplitudes and rates for an Unruh-DeWitt detector with dynamical internal mass-energy, thus incorporating a relativistic correction to the CoM, modeled in other contexts by the R{\" o}ntgen term or a mass defect. Our goal has been to extend a nonrelativistic model of the detector's CoM in a general manner, agnostic as to the details of the implementation of the detector. We have demonstrated that at least to lowest order perturbation in the interaction strength, the transition rates can be derived analytically.

The main result of this work is the necessity of including the relativistic corrections in the UDW model if any CoM dynamics is to be included at all. We carried out our derivations in the low-energy regime where relativistic effects have the least impact, yet, we found that mass-energy corrections are on the same order as the effects of the inclusion of the CoM dynamics itself. As a result there is no consistent  model of an UDW detector with fully nonrelativistic CoM---the mass-energy corrections are necessary.

Furthermore, there is the  ambiguity of choosing the value of the mass in the nonrelativistic limit, again relevant as the mass difference is also on the order of $E/m c^2$. We have argued that this choice should be made by defining the nonrelativistic limit as the one stemming from the interacting relativistic model in its low  energy regime (here low CoM momentum) whilst keeping a small but finite energy gap. We have found that this approach essentially identifies the nonrelativistic mass with the final mass-energy in the given process. That is, $M$ is identified with the ground-state mass $m_g$ in the emission case and the excited-state mass $m_e$ in the absorption case. 

One particular advantage of working in the low-energy regime is the potential transfer of the theory into the laboratory~\cite{Rosi2017,Loriani2019,DiPumpo2021}. 
Our result, and indeed already the existing literature, suggest that relativistic mass-energy corrections will again be necessary for consistent treatment of the detector recoil during the absorption or emission processes. As the UDW detector's recoil has been identified as a potential signature of the Unruh effect that could be experimentally observed~\cite{Sudhir2021}, it presents a natural context in which to implement our model and approach.
Apart from formal consistency of the results, the inclusion of the mass-energy equivalence also brings such a detector model closer to a composite particle, such as an atom, and thus also closer to the experimental realm.

From a theory perspective, inclusion of mass-energy equivalence in the UDW detector's dynamics results in a novel probe of joint gravitational and quantum phenomena---e.g.~the Hawking radiation surrounding a black hole. We expect this direction to extend a recent model of an UDW detector in a quantum-controlled superposition of classical trajectories~\cite{Barbado2020,Foo2020UDWSup}. The question of thermalization in our model is of particular interest, as such a system, with CoM and internal states coupled, would constitute a genuinely different model of a thermometer. 

There are also direct applications to relativistic quantum information, where particle detectors have long been a tool of choice, and entanglement ``harvesting'' with UDW detectors is a current area of particular interest. Already the impact of the inclusion of nonrelativistic CoM dynamics on the harvested entanglement, compared to classical models, has been shown~\cite{Stritzelberger2021}. It would be interesting to see whether the addition of the relativistic mass-energy effects in this context again leads to relevant corrections. Beyond the intuition arising from the present work, this may in particular be expected since the considered relativistic correction leads to internal states interacting, and so entangling, with the CoM. This suggests some interplay with the entanglement that these states can develop with the field.

\acknowledgments
We thank A.~Kempf and N.~Stritzelberger for early discussions, and also V.~Baccetti, J.~Foo, E.~Gale, L.~Henderson, and N.~Menicucci for further helpful discussions and advice.
M.Z.~acknowledges support through Australian Research Council (ARC) Future Fellowship grant No. FT210100675 and the ARC Centre of Excellence for Engineered Quantum Systems (EQUS), grant No. CE170100009. The authors acknowledge the traditional owners of the land on which the University of Queensland is situated.

\providecommand{\href}[2]{#2}\begingroup\raggedright\endgroup

\end{document}